\documentclass[twocolumn]{bmcart}

\usepackage{amsthm,amsmath,graphicx,gensymb,mathtools,geometry}

\usepackage[utf8]{inputenc} 
\usepackage[noadjust]{cite}

\usepackage[compact]{titlesec}  
\titlespacing{\section}{0pt}{20pt}{10pt}
\titlespacing{\subsection}{0pt}{20pt}{10pt}

\startlocaldefs
\definecolor{linkColor}{rgb}{1,0,0}

\newcommand{ \ii }{\emph{i} \,}
\newcommand{\tensor}[1]{\mathbf{#1}}

\endlocaldefs

\begin{document}

\begin{frontmatter}

\begin{fmbox}

\title{A Streaming Multi-GPU Implementation of Image Simulation Algorithms for Scanning Transmission Electron Microscopy}

\author[
   addressref={aff1},                  
   corref={aff1},                       
   email={apryor6@gmail.com}   
]{\inits{AP}\fnm{Alan} \snm{Pryor Jr.}}
\author[
   addressref={aff2},
   email={cophus@gmail.com}
]{\inits{CO}\fnm{Colin} \snm{Ophus}}
\author[
   addressref={aff1},
   email={miao@physics.ucla.edu}
    ]{\inits{JM}\fnm{Jianwei} \snm{Miao}}

\address[id=aff1]{
  \orgname{Department of Physics, University of California, Los Angeles}, 
  \street{Knudsen Hall, 475 Portola Plaza},                     %
  \postcode{90095},                                
  \city{Los Angeles},                              
  \cny{USA}                                    
}
\address[id=aff2]{                  
  \orgname{National Center for Electron Microscopy, Molecular Foundry, Lawrence Berkeley National Laboratory}, 
  \street{1 Cyclotron Road}, 
  \city{Berkeley},                              
  \cny{USA}                                    
}

{\small \emph{} \\ Alan Pryor Jr. 
\emph{apryor6@gmail.com} \\
Department of Physics, University of California, Los Angeles, CA, USA}

{\small \emph{} \\ Colin Ophus
\emph{cophus@gmail.com} \\
NCEM, Molecular Foundry, Lawrence Berkeley National Laboratory, Berkeley, CA, USA }

{\small \emph{} \\ Jianwei Miao
\emph{miao@physics.ucla.edu} \\
Department of Physics, University of California, Los Angeles, CA, USA}

\end{fmbox}


\begin{abstract} 
    Simulation of atomic resolution image formation in scanning transmission electron microscopy  can require significant computation times using traditional methods. A recently developed method, termed plane-wave reciprocal-space interpolated scattering matrix (PRISM), demonstrates potential for significant acceleration of such simulations with negligible loss of accuracy. Here we present a software package called \emph{Prismatic} for parallelized simulation of image formation in scanning transmission electron microscopy (STEM) using both the PRISM and multislice methods. By distributing the workload between multiple CUDA-enabled GPUs and multicore processors, accelerations as high as 1000x for PRISM and 15x for multislice are achieved relative to traditional multislice implementations using a single 4-GPU machine. We demonstrate a potentially important application of \emph{Prismatic}, using it to compute images for atomic electron tomography at sufficient speeds to include in the reconstruction pipeline. \emph{Prismatic} is freely available both as an open-source CUDA/C++ package with a graphical user interface and as a Python package, \emph{PyPrismatic}.
\end{abstract}


\end{frontmatter}

\section*{Introduction}

Scanning transmission electron microscopy (STEM) has had a major impact on materials science \cite{crewe1974scanning, nellist2007scanning}, especially for atomic-resolution imaging since the widespread adoption of hardware aberration correction \cite{batson2002sub, muller2009structure, pennycook2017impact}. Many large-scale STEM experimental techniques are routinely validated using imaging or diffraction simulations. Examples include electron ptychography \cite{pelz2017low}, 3D atomic reconstructions using dynamical scattering \cite{van2012method}, high precision surface atom position measurements on catalytic particles \cite{yankovich2014picometre}, de-noising routines \cite{mevenkamp2015poisson}, phase contrast imaging with phase plates \cite{ophus2016efficient}, new dynamical atomic contrast models \cite{van2016unscrambling}, atomic electron tomography (AET) \cite{miao_atomic_2016, xu2015three, yang2017deciphering, scott_electron_2012, chen_three-dimensional_2013}, and many others.  The most commonly employed simulation algorithm for STEM simulation is the multislice algorithm introduced by Cowlie and Moodie \cite{cowley1957scattering}. This method consists of two main steps. The first is calculation of the projected potentials from all atoms into a series of 2D slices. Second, the electron wave is initialized and propagated through the sample. The multislice method is straightforward to implement and is quite efficient for plane-wave or single-probe diffraction simulations \cite{kirkland1987simulation}.

\begin{table*}[t!]
\caption{A non-exhaustive list of electron microscopy simulation codes.}
\begin{center}
\begin{tabular}{lllll} 
    \hline
    Code(s) & Author(s) & Reference(s) & Comments & Links\\ 
    \hline
    \emph{xHREM} & Ishizuka  & \cite{ishizuka1977new, ishizuka2002practical} & 
    & \emph{https://www.hremresearch.com/Eng/simulation.html} 
    \\
    \emph{computem} & Kirkland & \cite{kirkland1987simulation,kirkland2010} & CPU  parallelized 
    & \emph{https://sourceforge.net/projects/computem/}
    \\
    \emph{EMS, JEMS} & Stadelmann & \cite{stadelmann1987ems, stadelmann2003image} & & \emph{http://www.jems-saas.ch/}
    \\
    \emph{MacTempas} & Kilaas & \cite{kilaas1990mactempas} & 
    & \emph{http://www.totalresolution.com/}
    \\
    \emph{QSTEM} & Koch & \cite{koch2002determination} & 
    & \emph{http://qstem.org/}
    \\
    \emph{CTEMsoft}   & De Graef & \cite{degraef2003introduction} & 
    & \emph{https://github.com/marcdegraef/CTEMsoft} 
    \\
    \emph{Web-EMAPS} & Zuo et al.\ & \cite{zuo2004web} & deprecated 
    & \emph{http://uiucwebemaps.web.engr.illinois.edu/}
    \\
    \emph{STEM\_CELL} & Carlino, Grillo et al.\ & \cite{carlino2008accurate, grillo2013stemcell} & CPU parallelized 
    & \emph{http://tem-s3.nano.cnr.it/?page\_id=2}
    \\
    \emph{STEMSIM} & Rosenauer and Schowalter & \cite{rosenauer2008stemsim} & 
    & \emph{http://www.ifp.uni-bremen.de/electron-microscopy/software/stemsim/}
    \\
    \emph{MALTS} & Walton et al.\ & \cite{walton2013malts} & Lorentz TEM 
    & 
    \\
    \emph{Dr.\ Probe} & Barthel and Houben & \cite{bar2012direct} & 
        & \emph{http://www.er-c.org/barthel/drprobe/}
    \\
    \emph{FDES} & Van den Broek et al.\ & \cite{van2015fdes} & GPU parallelized 
    & \emph{https://github.com/woutervandenbroek/FDES}
    \\ 
    \emph{$\mu$STEM} & D'Alfonso et al.\ & \cite{cosgriff2008three, forbes2010quantum} & GPU par., inelastic 
    & \emph{http://tcmp.ph.unimelb.edu.au/mustem/muSTEM.html}
    \\ 
    \emph{STEMsalabim} & Oelerich et al.\ & \cite{oelerich2017stemsalabim} & CPU parallelized 
    & \emph{http://www.online.uni-marburg.de/stemsalabim/}
    \\  
    \emph{Prismatic} & Pryor Jr.\ and Ophus & \cite{ophus2017fast}, this work & multi-GPU streaming 
    &  \emph{www.prism-em.com} and \emph{https://github.com/prism-em/prismatic}
    \\ 
    \hline
\end{tabular}
\end{center}
    \label{TableSTEMcodes}
\end{table*}

A large number of electron microscopy simulation codes are available, summarized in Table~\ref{TableSTEMcodes}. Most of these codes use the multislice method, and many have implemented parallel processing algorithms for both central processing units (CPUs) and graphics processing units (GPUs).  Recently some authors have begun using hybrid CPU+GPU codes for multislice simulation \cite{yao2016stem}. Multislice simulation relies heavily on the the fast Fourier transform (FFT) which can be computed using heavily optimized packages for both CPUs \cite{frigo2005design} and GPUs \cite{cufft}. The other primary computational requirement of multislice calculations is large element-wise matrix arithmetic, which GPUs are very well-suited to perform \cite{volkov2008benchmarking}. Parallelization is important because STEM experiments may record full probe images or integrated values from thousands or even millions of probe positions \cite{ophus2016efficient, yang2016simultaneous}. Performing STEM simulations on the same scale as these experiments is very challenging, because in the conventional multislice algorithm the propagation of each STEM probe through the sample is computed separately. Furthermore, if additional simulation parameters are explored the number of required simulations can become even larger, requiring very large computation times even using a modern, paralellized implementation. To address this issue, we introduced a new algorithm called PRISM which offers a substantial speed increase for STEM image simulations \cite{ophus2017fast}.

In this manuscript, we introduce a highly-optimized multi-GPU simulation code that can perform both multislice and PRISM simulations of extremely large structures called \emph{Prismatic}. We will briefly describe the multislice and PRISM algorithms, and describe the implementation details for our parallelized CPU and CPU+GPU codes. We perform timing benchmarks to compare both algorithms under a variety of conditions. Finally, we demonstrate the utility of our new code with typical use cases and compare with the popular package \emph{computem} \cite{kirkland2010}. \emph{Prismatic} includes a graphical user interface (GUI) and uses the cross-platform build system CMake \cite{martin2010mastering}. All of the source code is freely available.  Throughout this manuscript, we use the NVIDIA convention of referring to the CPU and GPU(s) as the host and device(s), respectively.


\section*{Methods}

\subsection*{Description of Algorithms}

\begin{figure*}[htbp!]
    \centering
        \includegraphics[width=5.75in]{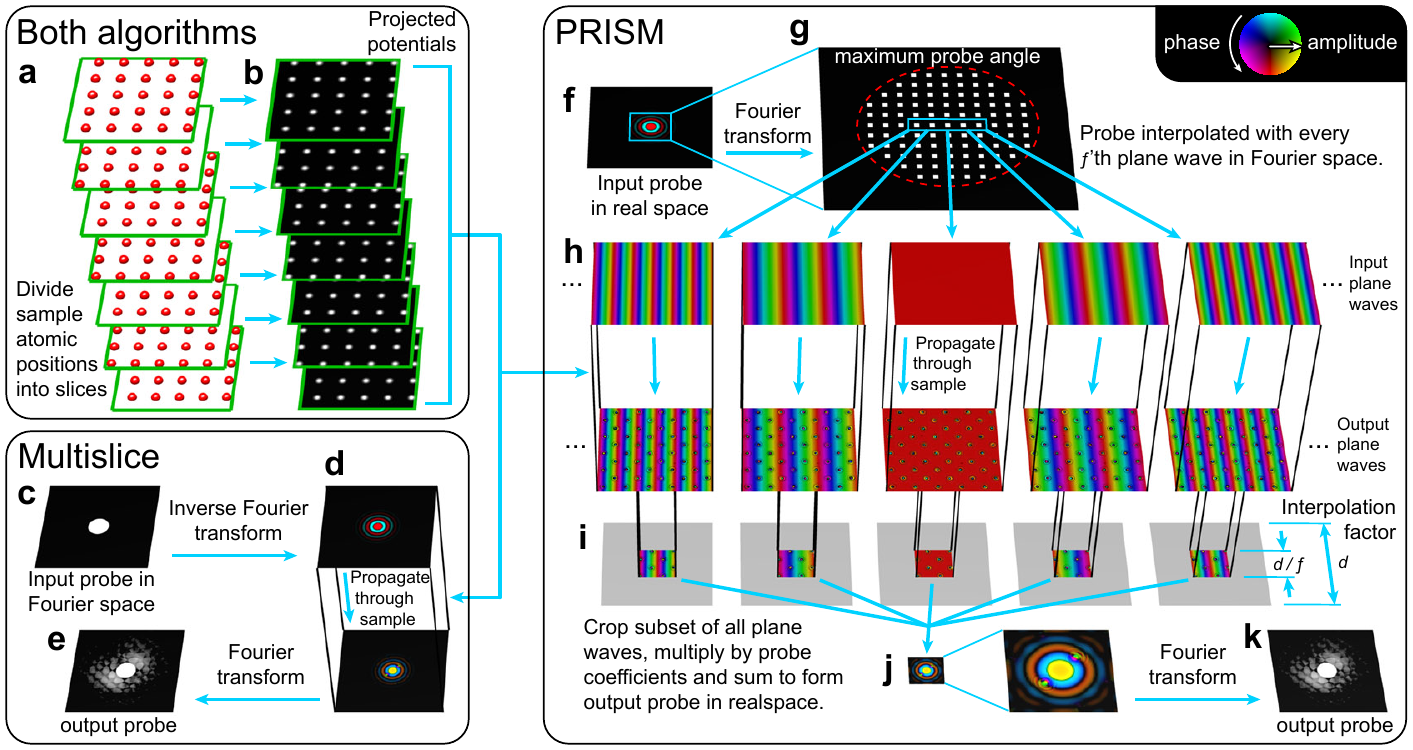}
 	\caption{Flow chart of STEM simulation algorithm steps. (a) All atoms are separated into slices at different positions along the beam direction, and (b) atomic scattering factors are used to compute projected potential of each slice. (c) Multislice algorithm, where each converged probe is initialized, (d) propagated through each of the sample slices defined in (b), and then (e) output either as images, or radially integrated detectors. (f) PRISM algorithm where (g) converged probes are defined in coordinate system downsampled by factor $f$ as a set of plane waves. (h) Each required plane wave is propagated through the sample slices defined in (b). (i) Output probes are computed by cropping subset of plane waves multiplied by probe complex coefficients, and (j) summed to form output probe, (k) which is then saved.}
	\label{FigureAlgorithms}
\end{figure*}

A flow chart of the steps performed in \emph{Prismatic} are given in Fig.~\ref{FigureAlgorithms}. Both multislice and PRISM share the same initial steps, where the sample is divided into slices which are used to compute the projected potential from the atomic scattering factors give in \cite{kirkland2010}. This step is shown schematically in Figs.~\ref{FigureAlgorithms}a and b, and is implemented by using a precomputed lookup table for each atom type \cite{ophus2016efficient, ophus2017fast}. 

Figs.\ref{FigureAlgorithms}c-e show the steps in a multislice STEM simulation. First the complex electron wave $\Psi$ representing the initial converged probe is defined, typically as an Airy disk function shown in Fig.~\ref{FigureAlgorithms}c. This probe is positioned at the desired location on the sample surface in realspace, as in Fig.~\ref{FigureAlgorithms}d. Next, this probe is propagated through the sample's potential slices defined in Fig.~\ref{FigureAlgorithms}b. This propagation is achieved by alternating two steps.  The first step is a transmission through a given potential slice $V_p^{\rm{2D}}$ over the realspace coordinates $\vec{r}$
\begin{equation}
    \psi_{p+1}(\vec{r}) =  \psi_p(\vec{r})
    \exp \left[ 
    \ii  \sigma V_p^{\rm{2D}}(\vec{r}) 
    \right],
\end{equation}
where $\sigma$ is the beam-sample interaction constant. Next, the electron wave is propagated over the distance $t$ to the next sample potential slice, which is done in Fourier space over the Fourier coordinates $\vec{q}$
\begin{equation}
    \Psi_{p+1}(\vec{q}) =  \Psi_p(\vec{q})
    \exp(- \ii \pi \lambda |\vec{q} \, |^2 t),
\end{equation}
where $\lambda$ is the electron wavelength. These steps are alternated until the electron probe has been propagated through the entire sample. Next, the simulated output is computed, which is typically a subset of the probe's intensity summed in Fourier space as shown in Fig.~\ref{FigureAlgorithms}e. The steps given in  Figs.~\ref{FigureAlgorithms}c-e are repeated for the desired probe positions, typically a 2D grid. The simulation result can be a single virtual detector, an array of annular ring virtual detectors or the entire probe diffraction pattern for each probe location, giving a 2D, 3D or 4D output respectively. For more details on the multislice method we refer readers to Kirkland \cite{kirkland2010}.

The PRISM simulation method for STEM images is outlined in Figs.~\ref{FigureAlgorithms}f-k. This method exploits the fact that an electron scattering simulation can be decomposed into an orthogonal basis set, as in the Bloch wave method \cite{kirkland2010}. If we compute the electron scattering for a set of plane waves that forms a complete basis, these waves can each be multiplied by a complex scalar value and summed to give a desired electron probe. A detailed description of the PRISM algorithm is given in \cite{ophus2017fast}.

The first step of PRISM is to compute the sample potential slices as in Figs.\ref{FigureAlgorithms}a-b. Next, a maximum input probe semi-angle and an interpolation factor $f$ is defined for the simulation. Fig.~\ref{FigureAlgorithms}g shows how these two variables specify the plane wave calculations required for PRISM, where every $f^{\rm{th}}$ plane wave in both spatial dimensions inside the maximum scattering angle is required. Each of these plane waves must be propagated through the sample using the mutlislice method given above, shown in Fig.~\ref{FigureAlgorithms}h. Once all of these plane waves have been propagated through the sample, together they form the desired basis set we refer to as the compact $\tensor{S}$-matrix. Next we define the location of all desired STEM probes. For each probe, a subset of all plane waves is cut out around the maximum value of the input STEM probe. The size length of the subset regions is $d/f$, where $d$ is the simulation cell length. The probe coefficients for all plane waves are complex values that define the center position of the STEM probe, and coherent wave aberrations such as defocus or spherical aberration. Each STEM probe is computed by multiplying each plane wave subset by the appropriate coefficient and summing all wave subsets. This is equivalent to using Fourier interpolation to approximate the electron probe wavefunction. As long as the subset region is large enough to encompass the vast majority of the probe intensity, the error in this approximation will be negligible \cite{ophus2017fast}. Finally, the output signal is computed for all probes as above, giving a 2D, 3D or 4D output array. As will be shown below, STEM simulations using the PRISM method can be significantly faster than using the multislice method.

\section*{Implementation Details}

\subsection*{Computational Model}

\begin{figure}[htbp!]
    \centering
        \includegraphics[width=2.7 in]{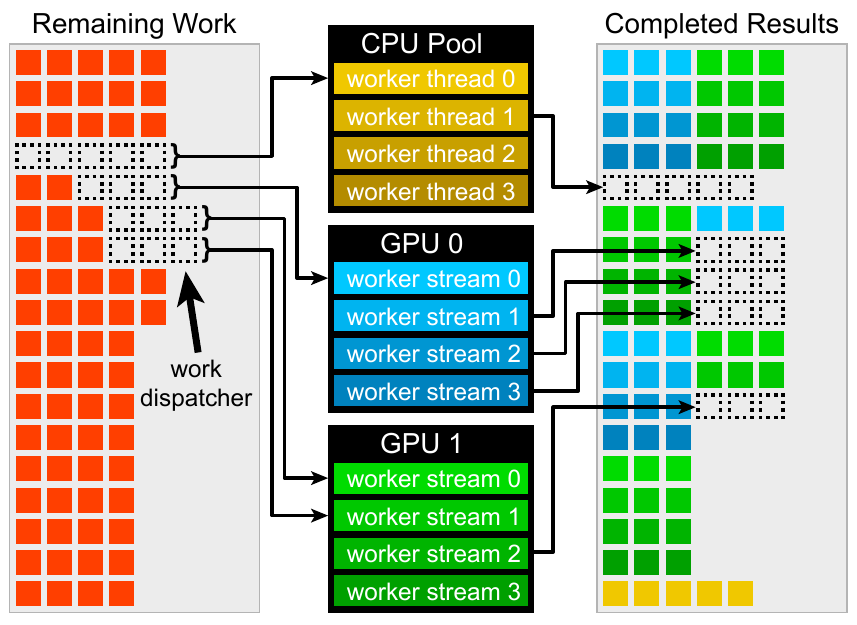}
         	\caption{Visualization of the computation model used repeatedly in the \emph{Prismatic} software package, whereby a pool of GPU and CPU workers are assigned batches of work by querying a synchronized work dispatcher. Once the assignment is complete, the worker requests more work until no more exists. All workers record completed simulation outputs in parallel.}
	\label{Figure_computationalModel}
\end{figure}

Wherever possible, parallelizable calculations in \emph{Prismatic} are divided into individual tasks and performed using a pool of CPU and GPU worker threads that asynchronously consume the work on the host or the device, respectively. We refer to a GPU worker thread as a host thread that manages work dispatched to a single device context. Whenever one of these worker threads is available, it queries a mutex-synchronized dispatcher that returns a unique work ID or range of IDs. The corresponding work is then consumed, and the dispatcher requeried until no more work remains. This computational model, depicted visually in Fig. \ref{Figure_computationalModel}, provides maximal load balancing at essentially no cost, as workers are free to independently obtain work as often as they become available. Therefore, machines with faster CPUs may observe more work being performed on the host, and if multiple GPU models are installed in the same system their relative performance is irrelevant to the efficiency of work dispatch. The GPU workers complete most types of tasks used by \emph{Prismatic} well over an order of magnitude faster than the CPU on modern hardware, and if a CPU worker is dispatched one of the last pieces of work then the entire program may be forced to unnecessarily wait on the slower worker to complete. Therefore, an adjustable early stopping mechanism is provided for the CPU workers.

\begin{figure}[htbp!]
    \centering
        \includegraphics[width=2.7in]{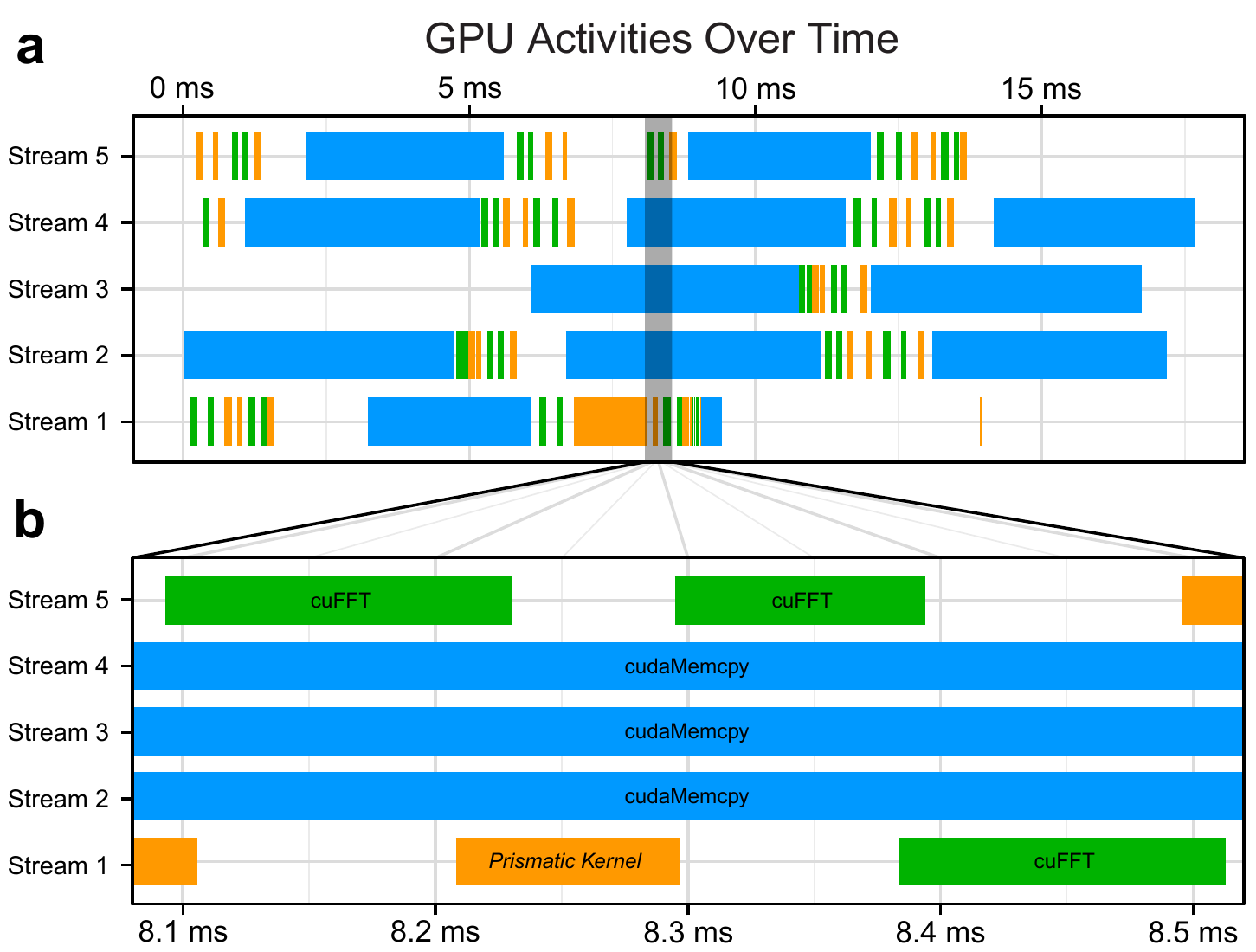}
        \caption{(a) Sample profile of the GPU activities on a single NVIDIA GTX 1070 during a multislice simulation in streaming mode with (b) enlarged inset containing a window where computation is occurring on streams \#1 and \#5 while three separate arrays are simultaneously being copied on streams \#2-4.}
	\label{FigureStreamingProfile}
\end{figure}

GPU calculations in \emph{Prismatic} are performed using a fully asynchronous memory transfer and computational model driven by CUDA streams. By default, kernel launches and calls to the CUDA runtime API for transferring memory occur on what is known as the default stream and subsequently execute in order. This serialization does not fully utilize the hardware, as it is possible to simultaneously perform a number of operations such as memory transfer from the host to the device, memory transfer from the device to the host, and kernel execution concurrently. This level of concurrency can be achieved using CUDA streams. Each CUDA stream represents an independent queue of tasks using a single device that execute internally in exact order, but that can be scheduled to run concurrently irrespective of other streams if certain conditions are met. This streaming model combined with the multithreaded work dispatch approach described previously allow for concurrent two-way host/device memory transfers and simultaneous data processing. A snapshot of the output produced by the NVIDA Visual Profiler for a single device context during a streaming multislice simulation similar to those described later in this work verifies that \emph{Prismatic} is indeed capable of such concurrency (Fig. \ref{FigureStreamingProfile}).

To achieve maximum overlap of work, each CUDA-enabled routine in \emph{Prismatic} begins with an initialization phase where relevant data on the host-side is copied into page-locked (also called ``pinned'') memory, which provides faster transfer times to the device and is necessary for asynchronous memory copying as the system can bypass internal staging steps that would be necessary for pageable memory \cite{cudaprogrammingguide}. CUDA streams and data buffers are then allocated on each device and copied to asynchronously. Read-only memory is allocated once per device, and read/write memory is allocated once per stream. It is important to perform all memory allocations initially, as any later calls to \emph{cudaMalloc} will implicitly force synchronization of the streams. Once the initialization phase is over, a host thread is spawned for each unique CUDA stream and begins to consume work.

\subsection*{Calculation of the Projected Potentials}

Both PRISM and multislice require dividing the atomic coordinates into thin slices and computing the projected potential for each. The calculation details are described by Kirkland and require evaluation of modified Bessel functions of the second kind, which are computationally expensive \cite{kirkland2010}. This barrier is overcome by precomputing the result for each unique atomic species and assembling a lookup table. Each projected potential is calculated on a supersampled grid, integrated, and cached. The sample volume is then divided into slices, and the projected potential for each slice is computed on separate CPU threads using the cached potentials. In principle this step could be GPU accelerated, but even for a large sample with several hundred thousand atoms the computation time is on the order of seconds and is considered negligible.

\subsection*{PRISM Probe Simulations}

Following calculation of the projected potential, the next step of PRISM is to compute the compact $\tensor{S}$-matrix. Each plane wave component is repeatedly transmitted and propagated through each slice of the potential until it has passed through the entire sample, at which point the complex-valued output wave is stored in real space to form a single layer of the compact $\tensor{S}$-matrix. This step of PRISM is highly analogous to multislice except whereas multislice requires propagating/transmitting the entire probe simultaneously, in PRISM each initial Fourier component is propagated/transmitted individually. The advantage is that in PRISM this calculation must only be performed once per Fourier component for the entire calculation, while in multislice it must be repeated entirely at every probe position. Thus, in many sample geometries the PRISM algorithm can significantly out-perform multislice despite the overhead of the $\tensor{S}$-matrix calculation \cite{ophus2017fast}.

The propagation step requires a convolution operation which can be performed efficiently through use of the FFT. Our implementation uses the popular FFTW and cuFFT libraries for the CPU and GPU implementations, respectively \cite{frigo2005design, cufft}. Both of these libraries support batch FFTs, whereby multiple Fourier transforms of the same size can be computed simultaneously. This allows for reuse of intermediate twiddle factors, resulting in a faster overall computation than performing individual transforms one-by-one at the expense of requiring a larger block of memory to hold the multiple arrays. \emph{Prismatic} uses this batch FFT method with both PRISM and multislice, and thus each worker thread will actually propagate a number of plane waves or probes simultaneously. This number, called the \emph{batch\_size}, may be tuned by the user to potentially enhance performance at the cost of using additional memory, but sensible defaults are provided.

In the final step of PRISM, a 2D output is produced for each probe position by applying coefficients, one for each plane wave, to the elements of the compact $\tensor{S}$-matrix and summing along the dimension corresponding to the different plane waves. These coefficients correspond to Fourier phase shifts that scale and translate each plane wave to the relevant location on the sample in real space. The phase coefficients, which are different for each plane wave but constant for a given probe position, are precomputed and stored in global memory. Each threadblock on the device first reads the coefficients from global memory into shared memory, where they can be reused throughout the lifetime of the threadblock. Components of the compact $\tensor{S}$-matrix for a given output wave position are then read from global memory, multiplied by the relevant coefficient, and stored in fast shared memory, where the remaining summation is performed. This parallel sum-reduction is performed using a number of well-established optimization techniques including reading multiple global values per thread, loop unrolling through template specialization, and foregoing of synchronization primitives when the calculation has been reduced to the single-warp level. Once the realspace exit wave has been computed, the modulus squared of its FFT yields the calculation result at the detector plane.

\subsection*{Multislice Probe Simulations}

The implementation of multislice is fairly straightforward. The initial probe is translated to the probe position of interest, and then is alternately transmitted and propagated through the sample. In practice this is accomplished by alternating forward and inverse Fourier transforms with an element-wise complex multiplication in between each with either the transmission or propagation functions. Upon propagation through the entire sample, the squared intensity of the Fourier transform of the exit wave provides the final result of the calculation at the detector plane for that probe position. For additional speed, the FFTs of many probes are computed simultaneously in batch mode. Thus in practice \emph{batch\_size} probes are transmitted, followed by a batch FFT, then propagated, followed by a batch inverse FFT, etc.

\subsection*{Streaming Data for Very Large Simulations}

The preferred way to perform PRISM and multislice simulations is to transfer large data structures such as the projected potential array or the compact $\tensor{S}$-matrix to each GPU only once, where they can then be read from repeatedly over the course of the calculation. However, this requires that the arrays fit into limited GPU memory. For simulations that are too large, we have implemented an asynchronous streaming version of both PRISM and multislice. Instead of allocating and transferring a single read-only copy of large arrays, buffers are allocated to each stream large enough to hold only the relevant subset of the data for the current step in the calculation, and the job itself triggers asynchronous streaming of the data it requires for the next step. For example, in the streaming implementation of multislice, each stream possesses a buffer to hold a single slice of the potential array, and after transmission through that slice the transfer of the next slice is requested. The use of asynchronous memory copies and CUDA streams permits the partial hiding of memory transfer latencies behind computation (Fig.~\ref{FigureStreamingProfile}). Periodically, an individual stream must wait on data transfer before it can continue, but if another stream is ready to perform work the device is effectively kept busy. Doing so is critical for performance, as the amount of time needed to transfer data can become significant relative to the total calculation. By default, \emph{Prismatic} uses an automatic setting to determine whether to use the single-transfer or streaming memory model whereby the input parameters are used to estimate how much memory will be consumed on the device, and if this estimate is too large compared with the available device memory then streaming mode is used. This estimation is conservative and is intended for convenience, but users can also forcibly set either memory mode. 

\subsection*{Launch Configuration}

All CUDA kernels are accompanied by a launch configuration that determines how the calculation will be carried out \cite{cudaprogrammingguide}. The launch configuration specifies the amount of shared memory needed, on which CUDA stream to execute the computation, and defines a 3D grid of threadblocks, each of which contains a 3D arrangement of CUDA threads. It is this arrangement of threads and threadblocks that must be managed in software to perform the overall calculation. The choice of launch configuration can have a significant impact on the overall performance of a CUDA application as certain GPU resources, such as shared memory, are limited. If too many resources are consumed by individual threadblocks, the total number of blocks that run concurrently can be negatively affected, reducing overall concurrency. This complexity of CUDA cannot be overlooked in a performance-critical application, and we found that the speed difference in a suboptimal and well-tuned launch configuration could be as much as 2-3x.

In the reduction step of PRISM, there are several competing factors that must be considered when choosing a launch configuration. The first of these is the threadblock size. The compact $\tensor{S}$-matrix is arranged in memory such that the fastest changing dimension, considered to be the x-axis, lies along the direction of the different plane waves. Therefore to maximize memory coalescence, threadblocks are chosen to be as large as possible in the x-direction. Usually the result will be threadblocks that are effectively 1D, with $BlockSize_y$ and $BlockSize_z$ equal to one; however in cases where very few plane waves need to be computed the blocks may be extended in y and z to prevent underutilization of the device. To perform the reduction, two arrays of shared memory are used. The first is dynamically sized and contains as many elements as there are plane waves. This array is used to cache the phase shift coefficients to prevent unnecessary reads from global memory, which are slow. The second array has $BlockSize_x$*$BlockSize_y$*$BlockSize_z$ elements and is where the actual reduction is performed. Each block of threads steps through the array of phase shifts once and reads them into shared memory. Then the block contiguously steps through the elements of the compact $\tensor{S}$-matrix for a different exit-wave position at each y and z index, reading values from global memory, multiplying them by the associated coefficient, and accumulating them in the second shared memory array. Once all of the plane waves have been accessed, the remaining reduction occurs quickly as all remaining operations occur in fast shared memory. Each block of threads will repeat this process for many exit-wave positions which allows efficient reuse of the phase coefficients from shared memory. The parallel reduction is performed by repeatedly splitting each array in half and adding one half to the other until only one value remains. Consequently, if the launch configuration specifies too many threads along the x-direction, then many of them will become idle as the reduction proceeds, which wastes work. Conversely, choosing $BlockSize_x$ to be too small is problematic for shared memory usage, as the amount of shared memory per block for the phase coefficients is constant regardless of the block size. In this case, the amount of shared memory available will rapidly become the limiting factor to the achievable occupancy. A suitably balanced block size produces the best results.

The second critical component of the launch configuration is the number of blocks to launch. Each block globally reads the phase coefficients once and then reuses them, which favors using fewer blocks and having each compute more exit-wave positions. However, if too few blocks are launched the device may not reach full occupancy. The theoretically optimal solution would be to launch the minimal amount of blocks needed to saturate the device and no more.

 Considering these many factors, \emph{Prismatic} uses the following heuristic to choose a good launch configuration. At runtime, the properties of the available devices are queried, which includes the maximum number of threads per threadblock, the total amount of shared memory, and the total number of streaming multiprocessors. $BlockSize_x$ is chosen to be either the largest power of two smaller than the number of plane waves or the maximum number of threads per block, whichever is smaller. The total number of threadblocks that can run concurrently on a single streaming multiprocessor is then estimated using $BlockSize_x$, the limiting number of threads per block, and the limiting number of threadblocks per streaming multiprocessor.  The total number of threadblocks across the entire device is then estimated as this number times the total number of streaming multiprocessors, and then the grid dimensions of the launch configuration are set to create three times this many blocks, where the factor of three is a fudge factor that we found produces better results.  

\section*{Benchmarks}

\subsection*{Algorithm Comparison}

\begin{figure*}[htbp!]
    \includegraphics[width=5.0in]{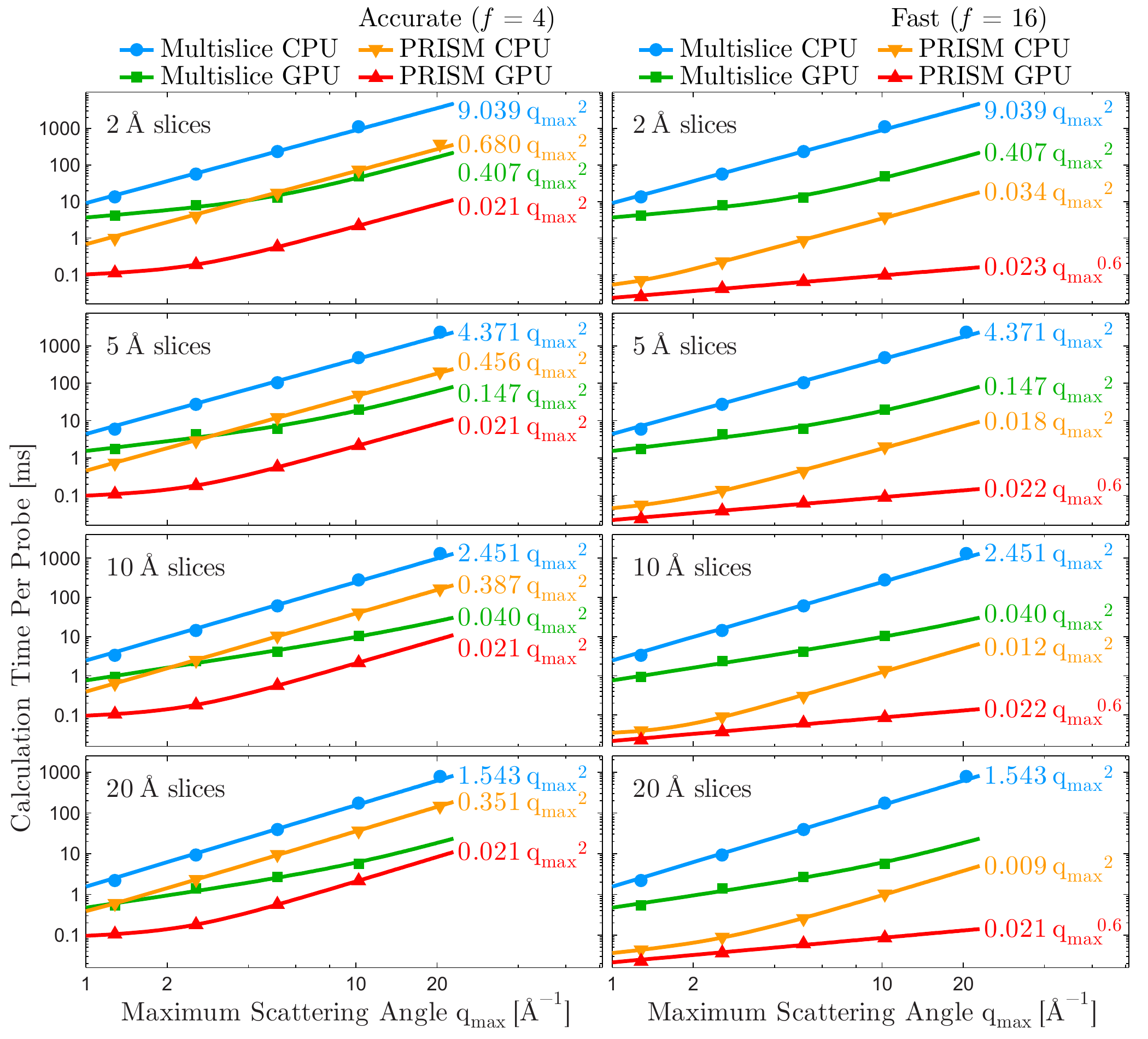}
 	\caption{Comparison of the CPU/GPU implementations of the PRISM and multislice algorithms described in this work.  A 100x100x100 $\rm{\AA}$ amorphous carbon cell was divided slices of varying thickness and sampled with increasingly small pixels in real space corresponding to digitized probes of array size 256x256, 512x512, 1024x1024, and 2048x2048, respectively. Two different PRISM simulations are shown, a more accurate case where the interpolation factor $f=4$ (left), and a faster case with $f=16$ (right). The multislice simulation is the same for both columns. Power laws were fit of the form $A + B \, {q_{\rm{max}}}^n$  where possible. The asymptotic power laws for higher scattering angles are shown on the right of each curve.}
	\label{FigureAbsoluteBenchmark}
\end{figure*}

A total of four primary algorithms are implemented \emph{Prismatic}, as there are optimized CPU and GPU implementations of both PRISM and multislice simulation. To visualize the performance of the different algorithms, we performed a number of benchmarking simulations spanning a range of sample thicknesses, sizes, and with varying degrees of sampling. Using the average density of amorphous carbon, an atomic model corresponding to a 100x100x100 $\rm{\AA}$  carbon cell was constructed and used for image simulation with various settings for slice thickness and pixel sampling. The results of this analysis are summarized in Fig.~\ref{FigureAbsoluteBenchmark}. These benchmarks are plotted as a function of the maximum scattering angle  $q_{\rm{max}}$, which varies inversely to the pixel size.

The difference in computation time $t$ shown in Fig.~\ref{FigureAbsoluteBenchmark} between traditional CPU multislice and GPU PRISM is stark, approximately four orders of magnitude for the ``fast'' setting where $f=16$, and still more than a factor of 500 for the more accurate case of $f=4$. For both PRISM and multislice, the addition of GPU acceleration increases speed by at least an order of magnitude. Note that as the thickness of the slices is decreased, the relative gap between PRISM and multislice grows, as probe calculation in PRISM does not require additional propagation through the sample.  We have also fit trendline curves of the form
\begin{equation}
    t = A + B \, {q_{\rm{max}}}^n,
\end{equation}
where $A$ and $B$ are prefactors and $n$ is the asymptotic power law for high scattering angles.  We observed that most of the simulation types approximately approach $n=2$, which is unsurprising for both PRISM and multislice. The limiting operation in PRISM is matrix-scalar multiplication,  which depends on the array size and varies as ${q_{\rm{max}}}^2$. For multislice the computation is a combination of multiplication operations and FFTs, and the theoretical  $\mathcal{O}(n\log{}n)$ scaling of the latter is only slightly larger than 2, and thus the trendline is an approximate lower bound. The only cases that fall significantly outside the $n=2$ regime were the multislice GPU simulations with the largest slice separation (20 $\rm{\AA}$) and the ``fast'' PRISM GPU simulations where $f=16$. These calculations are sufficiently fast that the relatively small overhead required to compute the projected potential slices, allocate data, etc., is actually a significant portion of the calculation, resulting in scaling better than ${q_{\rm{max}}}^2$. For the $f=16$ PRISM case, we observed approximately ${q_{\rm{max}}}^{0.6}$ scaling, which translates into sub-millisecond calculation times per probe even with small pixel sizes and slice thicknesses.


To avoid unnecessarily long computation times for the many simulations, particularly multislice, different numbers of probe positions were calculated for each algorithm, and thus we report the benchmark as time per probe. Provided enough probe positions are calculated to obviate overhead of computing the projected potential and setting up the remainder of the calculation, there is a linear relationship between the number of probe positions calculated and the calculation time for all of the algorithms, and computing more probes will not change the time per probe significantly. Here this overhead is only on the order of 10 seconds or fewer, and the reported results were obtained by computing 128x128 probes for PRISM CPU and multislice CPU, 512x512 for multislice GPU, and 2048x2048 for PRISM GPU. All of these calculations used the single-transfer memory implementations and were run on compute nodes with dual Intel Xeon E5-2650 processors, four Tesla K20 GPUs, and 64GB RAM  from the VULCAN cluster within the Lawrence Berkeley National Laboratory Supercluster.

\subsection*{Hardware Scaling}

Modern high performance computing is dominated by parallelization. At the time of this writing virtually all desktop CPUs contain at least four cores, and high end server CPUs can have as many as twenty or more. Even mobile phones have begun to routinely ship with multicore processors \cite{sakran2007implementation}. In addition to powerful CPUs, GPUs and other types of coprocessors such as the Xeon Phi \cite{Jeffers:2013:IXP:2523262} can be used to accelerate parallel algorithms. It therefore is becoming increasingly important to write parallel software that fully utilizes the available computing resources.

\begin{figure}[htbp!]
    \includegraphics[width=2.7in]{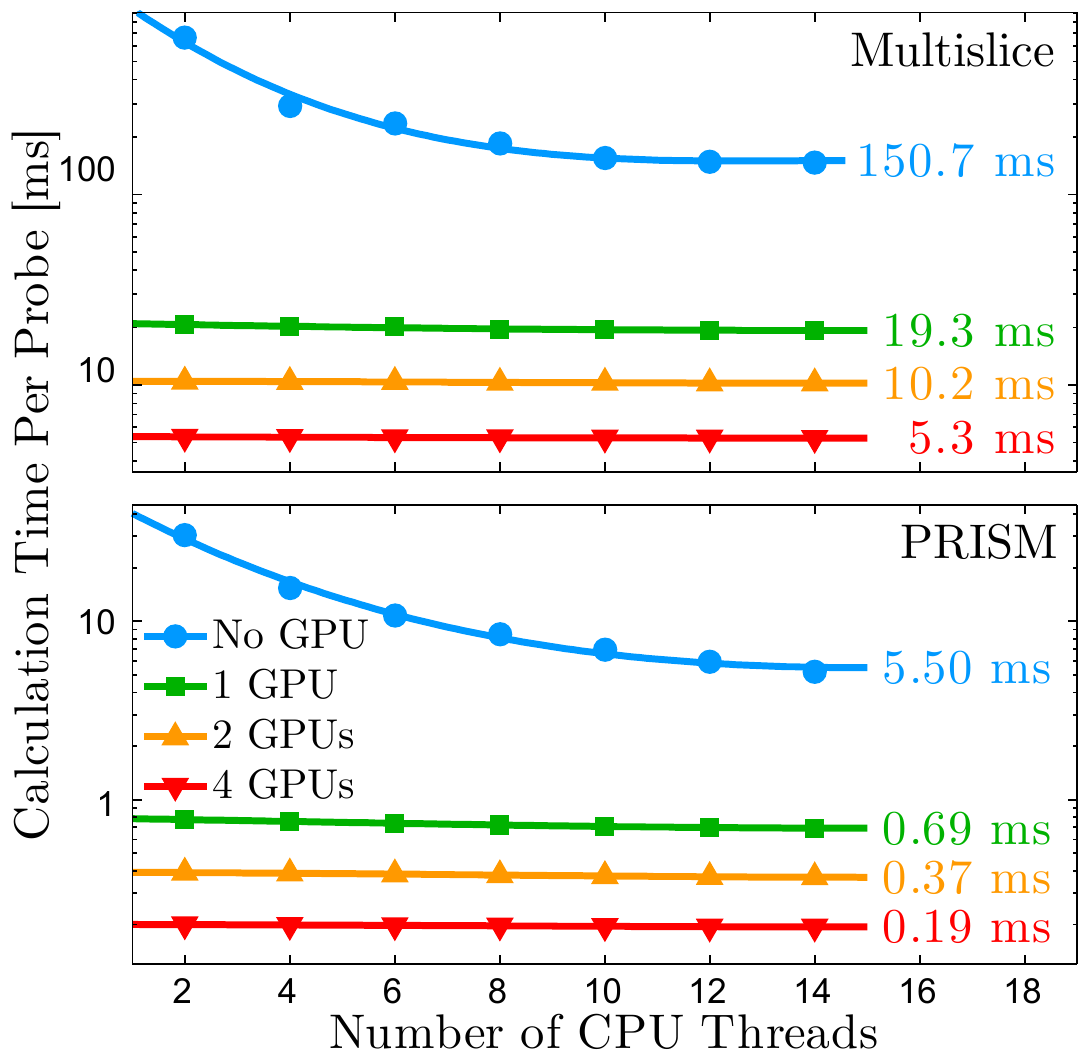}
 	\caption{Comparison of the implementations of multislice and PRISM for varying combinations of CPU threads and GPUs. The simulation was performed on a 100x100x100 $\rm{\AA}$ amorphous carbon cell with 5 $\rm{\AA}$ thick slices, and 0.1 $\rm{\AA}$ pixel size. All simulations were performed on compute nodes with dual Intel Xeon E5-2650 processors, four Tesla K20 GPUs, and 64GB RAM. Calculation time of rightmost data point is labeled for all curves.}
	\label{FigureHardwareBenchmark}
\end{figure}

To demonstrate how the algorithms implemented in \emph{Prismatic} scale with hardware, we performed the following simulation. Simulated images of a 100x100x100 $\rm{\AA}$ amorphous carbon cell were produced with both PRISM and multislice using 5 $\rm{\AA}$ thick slices, pixel size 0.1 $\rm{\AA}$, and 80 keV electrons. This simulation was repeated using varying numbers of CPU threads and GPUs. As before, a varying number of probes was computed for each algorithm, specifically 2048x2048 for GPU PRISM, 512x512 for CPU PRISM and GPU multislice, and 64x64 for CPU multislice. This simulation utilized the same 4-GPU VULCAN nodes described previously. The results of this simulation are summarized in Fig.~\ref{FigureHardwareBenchmark}.

The ideal behavior for the CPU-only codes would be to scale as $1/x$ with the number of CPU cores utilized such that doubling the number of cores also approximately doubles the calculation speed. Provided that the number of CPU threads spawned is not greater than the number of cores, the number of CPU threads can effectively be considered the number of CPU cores utilized, and this benchmark indicates that both CPU-only PRISM and multislice possess close to ideal scaling behavior with number of CPU cores available.

The addition of a single GPU improves both algorithms by approximately a factor of 8 in this case, but in general the relative improvement varies depending on the quality and number of the CPUs vs GPUs. The addition of a second GPU improves the calculation speed by a further factor of 1.8-1.9 with 14 threads, and doubling the number of GPUs to four total improves again by a similar factor. The reason that this factor is less than two is because the CPU is doing a nontrivial amount of work alongside the GPU. This claim is supported by the observation that when only using two threads the relative performance increase is almost exactly a factor of two when doubling the number of GPUs. We conclude that our implementations of both algorithms scale very well with available hardware, and potential users should be confident that investing in additional hardware, particularly GPUs, will benefit them accordingly.

\subsection*{Data Streaming/Single-Transfer Benchmark}

\begin{figure}[htbp!]
    \centering
        \includegraphics[width=2.7in]{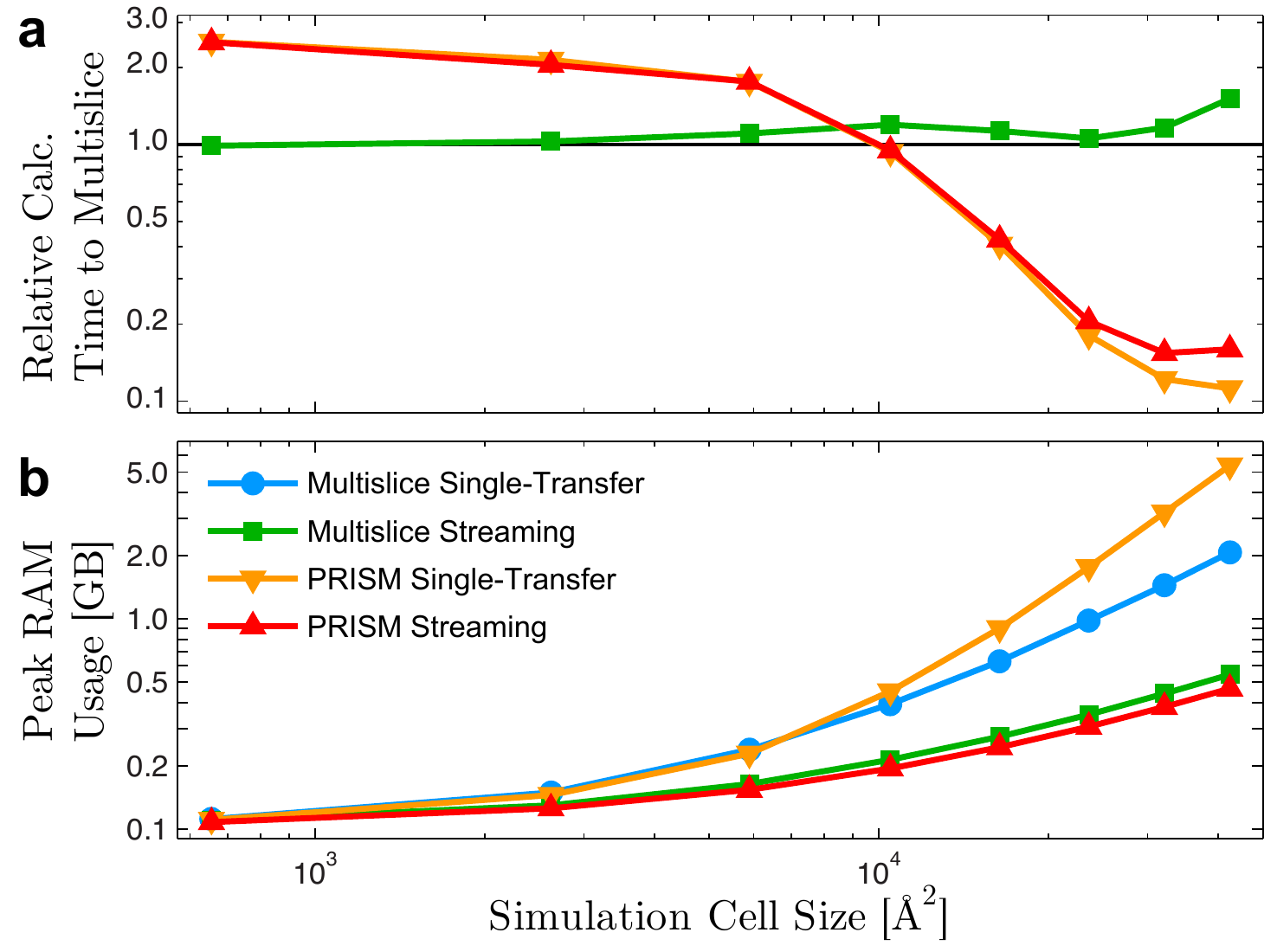}
         	\caption{Comparison of (a) relative performance and (b) peak memory consumption for single transfer and streaming implementations of PRISM and multislice.}
	\label{FigureStreamingBenchmark}
\end{figure}

 For both PRISM and multislice, \emph{Prismatic} implements two different memory models, a single-transfer method where all data is copied to the GPU a single time before the main computation begins, and a streaming mode where asynchronous copying of the required data is triggered across multiple CUDA streams as it is needed throughout the computation. Streaming mode reduces the peak memory required on the device at the cost of redundant copies; however, the computational cost of this extra copying can be partially alleviated by hiding the transfer latency behind compute kernels and other copies (Fig. \ref{FigureStreamingProfile}).
 To compare the implementations of these two memory models in \emph{Prismatic}, a number of amorphous carbon cells of increasing sizes were used as input to simulations using 80 keV electrons, 20 mrad probe convergence semi-angle, 0.1 $\rm{\AA}$ pixel size, 4 $\rm{\AA}$ slice thickness, and 0.4 $\rm{\AA}$ probe steps.  Across a range of simulation cell sizes the computation time of the streaming vs. single-transfer versions of each code are extremely similar while the peak memory may be reduced by an order of magnitude or more (Fig. \ref{FigureStreamingBenchmark}). For the streaming calculations, memory copy operations may become significant relative to the computational work (Fig. \ref{FigureStreamingProfile});however, this can be alleviated by achieving multi-stream concurrency.

\subsection*{Comparison to existing methods}

All previous benchmarks in this work have measured the speed of the various algorithms included in \emph{Prismatic} against each other; however, relative metrics are largely meaningless without an external reference both in terms of overall speed and resulting image quality. To this end, we also performed STEM simulations of significant size and compare the results produced by the algorithms in \emph{Prismatic} and the popular package \emph{computem} \cite{kirkland1987simulation, kirkland2010}. 

We have chosen a simulation cell typical of those used in structural atomic-resolution STEM studies, a complex Ruddlesden–Popper (RP) layered oxide.  The RP structure we used contains 9 pseudo-cubic unit cells of perovskite strontium titanate structure, with two stacking defects every 4.5 1x1 cells that modify the composition and atomic coordinates. The atomic coordinates of this cell were refined using Density Functional Theory and were used for very-large-scale STEM image simulations \cite{stone2016atomic}. This 9x1x1 unit cell was tiled 4x36x25 times resulting in final sample approximately 14 x 14 nm in-plane and 10 nm thick, containing roughly 1.4 million atoms. 

\begin{figure*}[htbp!]
    \centering
        \includegraphics[width=5.5in]{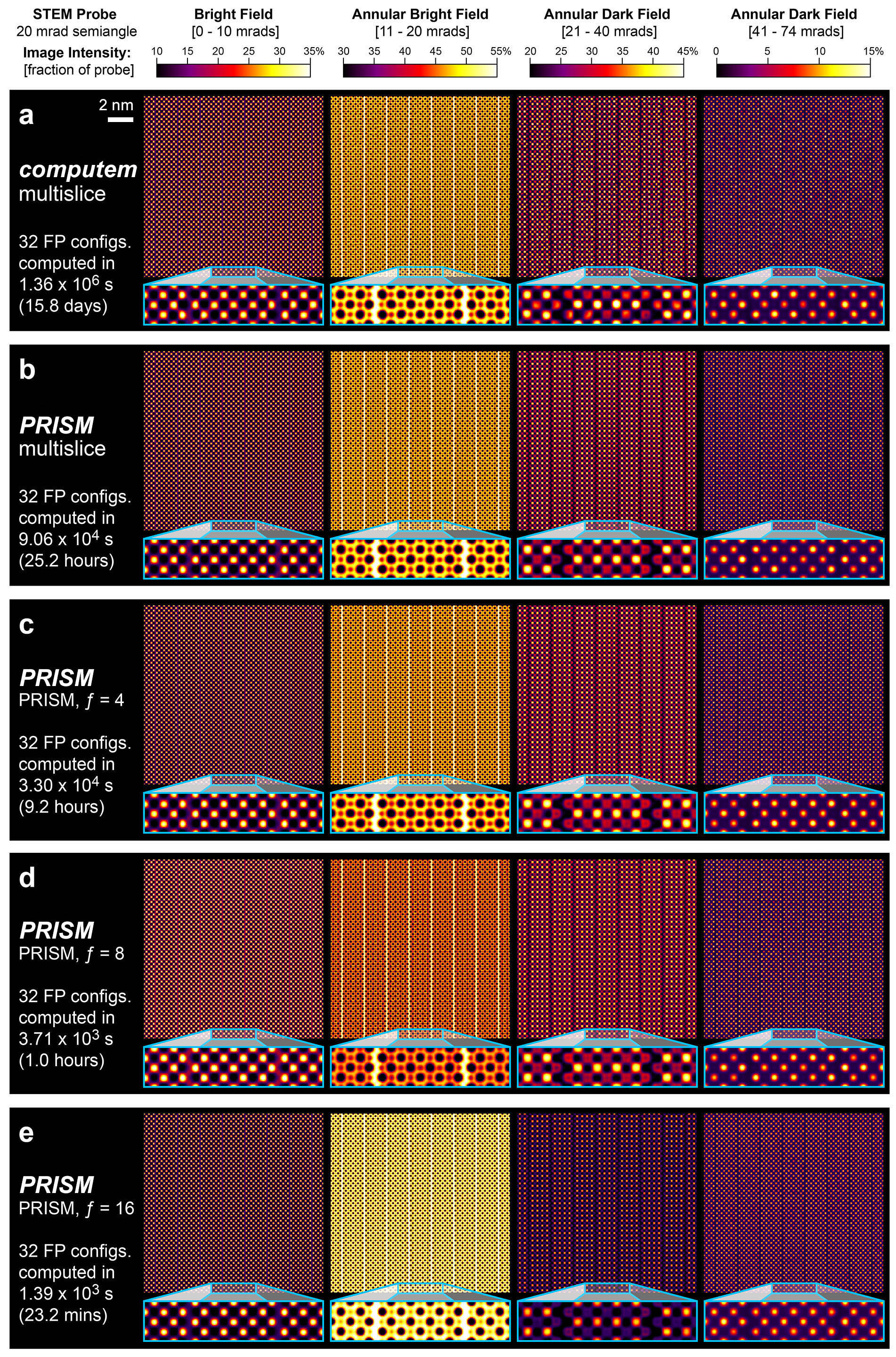}
         	\caption{Comparison of simulation results produced by \emph{computem} and \emph{Prismatic}. The sample is composed of  36x36x25 pseudocubic perovskite unit cells, and images were simulated using 80 keV electrons, a 20 mrad probe convergence semi-angle, 0 $\rm{\AA}$ defocus, and 1024x1024 pixel sampling for the probe and projected potential. A total of 720x720 probe positions were computed and the final images are an average over 32 frozen phonon configurations. Separate PRISM simulations were performed with interpolation factors 4, 8, and 16.}
	\label{FigureKirklandComparison}
\end{figure*}

Simulations were performed with multislice as implemented in \emph{computem} (specifically using the \emph{autostem} module), multislice in \emph{Prismatic}, and the PRISM method with $f$ values of 4, 8 and 16 using 80 keV electrons, 1024 x 1024 pixel sampling, 20 mrad probe convergence semi-angle, and 5 $\rm{\AA}$ thick potential slices. A total of 720x720 evenly spaced probes were computed for each simulation, and a total of 32 frozen phonon configurations were averaged to produce the final images, which are summarized in Fig.~\ref{FigureKirklandComparison}. The PRISM algorithms were run on the VULCAN GPU nodes while \emph{computem} simulations utilized better VULCAN CPU nodes with dual Intel Xeon E5-2670v2 CPUs and 64GB RAM. 

The mean computation time per frozen phonon for the \emph{computem} simulations was 709.8 minutes resulting in a total computation time of 15.8 days. The use of our GPU multislice code here provides an acceleration of about 15x, reducing the computation from more than two weeks to just over one day. The PRISM $f=4$ simulation is almost indistinguishable from the multislice results, and gives a 2.7x speed up over our GPU multislice simulation. For the $f=8$ PRISM simulation, some intensity differences are visible in the two bright field images, but the relative contrast of all atomic sites is still correct. This simulation required just over an hour, providing a speedup of 25X relative to our GPU multislice simulation. The $f=16$ PRISM result show substantial intensity deviations from the ideal result, but require just 43 seconds per frozen phonon configuration. The total difference in acceleration from CPU multislice to the fastest PRISM simulation shown in Fig.~\ref{FigureKirklandComparison} is just under three orders of magnitude. Ultimately, the user’s purpose dictates what balance of speed and accuracy is appropriate, but the important point is that calculations that previously required days or weeks on a computer cluster may now be performed on a single workstation in a fraction of the time.

\section*{Application to Atomic Electron Tomography}

\begin{figure*}[htbp!]
    \includegraphics[width=5.0in]{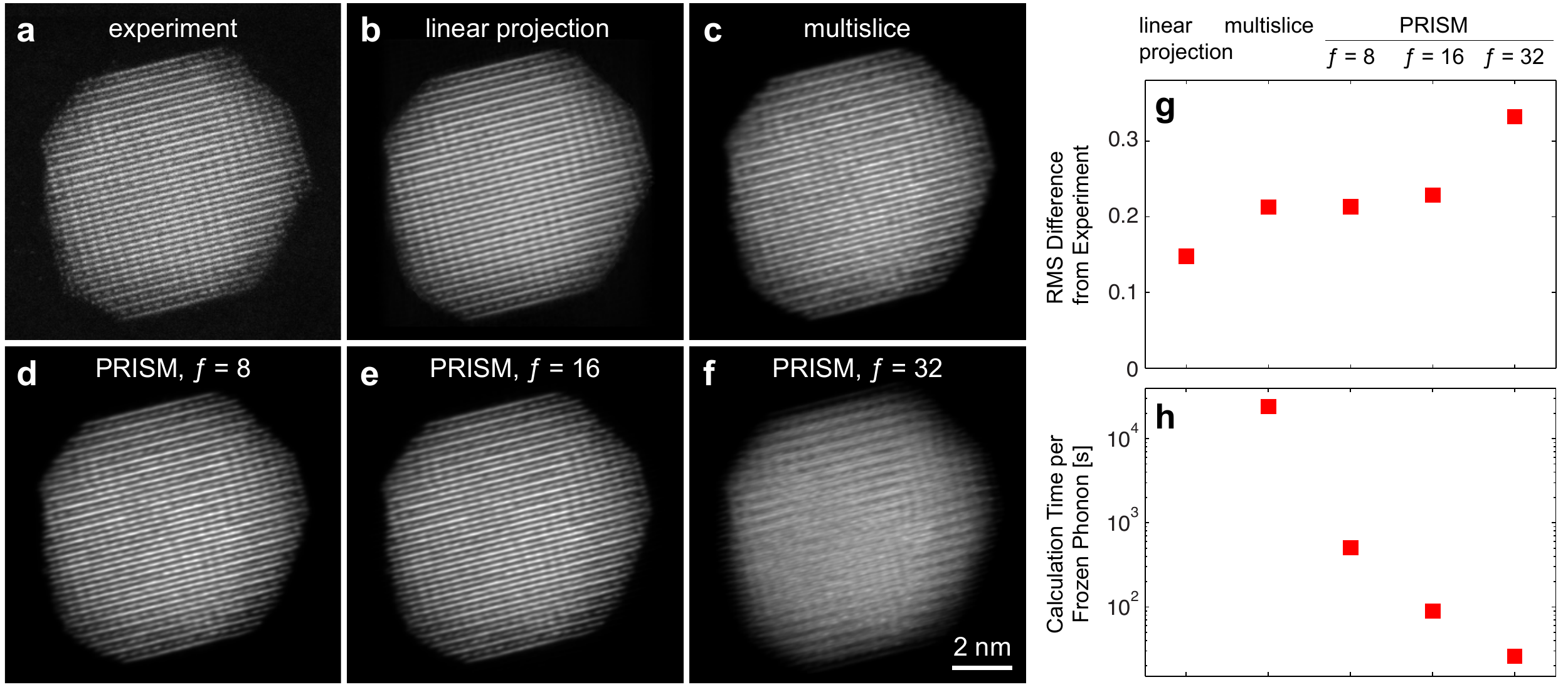}
  \caption{Images from one projection of an atomic electron tomography tilt series of an FePt nanoparticle \cite{yang2017deciphering}, from (a) experiment, (b) linear projection of the reconstruction, (c) multislice simulation, and (d)-(f) PRISM simulations for $f=8$, 16, and 32 respectively. (g) Relative root-mean-square error of the images in (b)-(f) relative to (a). (h) Calculation times per frozen phonon configuration for (c)-(f). All simulations performed with \emph{Prismatic}.}
  \label{FigureApplication}
\end{figure*}

One potentially important application of STEM image simulations is AET experiments. One of the ADF-STEM images from an atomic resolution tilt series of an FePt nanoparticle \cite{yang2017deciphering} is shown in Fig.~\ref{FigureApplication}a, with the corresponding linear projection from the 3D reconstruction shown in Fig.~\ref{FigureApplication}b. In this study and others, we have used multislice simulations to validate the tomographic reconstructions and estimate both the position and chemical identification errors \cite{xu2015three, yang2017deciphering}. One such multislice simulation is given in Fig.~\ref{FigureApplication}c. This simulation was performed at 300 kV using a 30 mrad STEM probe, with a simulation pixel size of 0.0619 $\rm{\AA}$ and a spacing between adjacent probes of 0.3725 $\rm{\AA}$. The image results shown are for 16 frozen phonon configurations using a 41-251 mrad annular dark field detector. This experimental dataset includes some postprocessing and was obtained freely online \cite{yang2017deciphering}.

The 3D reconstruction algorithm we have used, termed GENeralized Fourier Iterative REconstruction (GENFIRE), assumes that the projection images are linearly related to the potential of the reconstruction \cite{yang2017deciphering, pryor2017genfire}. This assumption was sufficient for atomic resolution tomographic reconstruction, but the measured intensity has some non-linear dependence on the atomic potentials, due to effects such as exponential decrease of electrons in the unscattered STEM probe, channeling effects along atomic columns, coherent diffraction at low scattering angles and other related effects \cite{muller2004atomic, lebeau2008quantitative, findlay2010dynamics, kourkoutis2011direct, van2016unscrambling, woehl2016dark, cui2017origin}. These effects can be seen in the differences between the images shown in Figs.~\ref{FigureApplication}b and c. The multislice simulation image shows sharper atomic columns, likely due to the channeling effect along atomic columns that are aligned close to the beam direction \cite{findlay2010dynamics}. Additionally, there are mean intensity differences between the center part of the the particle (thickest region) and the regions closed to the surfaces in projection (thinnest regions). Including these dynamical scattering effects in the reconstruction algorithm would increase the accuracy of the reconstruction.

However, Fig.~\ref{FigureApplication}h shows that the computation time for the multislice simulation is prohibitively high. Even using the \emph{Prismatic} GPU code, each frozen phonon configuration for multislice require almost 7 hours. Using 16 configurations and simulating all 65 projection angles would require months of simulation time, or massively parallel simulation on a super cluster. An alternative is to use the PRISM algorithm for the image simulations, shown in Figs.~\ref{FigureApplication}d, e and f for interpolation factors of $f=8$, $16$ and $32$ respectively. Fig.~\ref{FigureApplication}g shows the relative errors of Figs.~\ref{FigureApplication}b-f, where the error is defined by the root-mean-square of the intensity difference with the experimental image in Fig.~\ref{FigureApplication}a, divided by the root-mean-square of the experimental image. Unsurprisingly, the linear projection shows the lowest error since it was calculated directly from the 3D reconstruction built using the experimental data. The multislice and PRISM $f=8$ and $f=16$ simulations show essentially the same errors within the noise level of the experiment. The PRISM $f=32$ has a higher error, and obvious image artifacts are visible in Figs.~\ref{FigureApplication}f. Thus, we conclude that using an interpolation factor $f=16$ produces an image of sufficient accuracy. This calculation required only 90 s per frozen phonon calculation, and therefore computing 16 configuration for all 65 tilt angles would require only 26 hours. One could therefore imagine integrating this simulation routine into the final few tomography reconstruction iterations to account for dynamical scattering effects and to improve the reconstruction quality.

\section*{Conclusion}

We have presented \emph{Prismatic}, an asynchronous, streaming multi-GPU implementation of the PRISM and multislice algorithms for image formation in scanning transmission electron microscopy. Both multislice and PRISM algorithms were described in detail as well as our approach to implementing them in a parallel framework. Our benchmarks demonstrate that this software may be used to simulate STEM images up to several orders of magnitude faster than using traditional methods, allowing users to simulate complex systems on a GPU workstation without the need for a computer cluster. \emph{Prismatic} is freely available as an open-source C++/CUDA package with a graphical interface that contains convenience features such as allowing users to interactively view the projected potential slices, compute/compare individual probe positions with both PRISM and multislice, and dynamically adjust positions of virtual detectors. A command line interface and a Python package, \emph{PyPrismatic}, are also available. We have demonstrated one potential application of the \emph{Prismatic} code, using it to compute STEM images to improve the accuracy in atomic electron tomography. We hope that the speed of this code as well as the convenience of the user interface will have significant impact for users in the EM community.

\begin{backmatter}

\section*{Competing interests}
  The authors declare that they have no competing interests.

\section*{Author's contributions}
    AP designed the software, implemented the CUDA/C++ versions of PRISM and multislice, programmed the graphical user interface and command line interface, and performed the simulations in this paper. CO conceived of the PRISM algorithm, wrote the original MATLAB implementations, and made the figures. AP and CO rote the manuscript. JM advised the project. All authors commented on the manuscript.

\section*{Acknowledgements}

The computations were supported by a User Project at The Molecular Foundry using its compute cluster (VULCAN), managed by the High Performance Computing Services Group, at Lawrence Berkeley National Laboratory (LBNL), and supported by the Office of Science of the U.S. Department of Energy under contract No. DE-AC02-05CH11231.

\section*{Data availability}
The \emph{Prismatic} source code, installers, and documentation with tutorials are freely available at www.prism-em.com

\bibliographystyle{bmc-mathphys}
\bibliography{refs}    

\end{backmatter}
\end{document}